\documentclass[twocolumn,showpacs,preprintnumbers,amsmath,amssymb]{revtex4}

\usepackage{graphicx}
\usepackage{dcolumn}
\usepackage{bm}
\usepackage{color}
\usepackage{graphicx}
\usepackage{epstopdf}
\usepackage[hang,scriptsize]{subfigure}

\begin{document}

\title{Critical Points of Correlated Percolation in a
Gravitational Link-adding Network Model}

\author{Chen-Ping Zhu}
\affiliation{College of Science, Nanjing University of Aeronautics
and Astronautics, Nanjing, 210016, China}
\affiliation{Max-Plank-Institute f$\ddot{u}$r Physik Komplexer
Systeme, N$\ddot{o}$thnitzer Stra$\ss$e, 38, 01187 Dresden, Germany}
\affiliation{%
Research center for complex system science, Shanghai University for
Science and Technology, 200093, China
}%
\affiliation{%
Kavli Institute for Theoretic Physics, Beijing, 100093, China
}%
\author{Long-Tao Jia}
\affiliation{College of Science, Nanjing University of Aeronautics
and Astronautics, Nanjing, 210016, China}
\author{Beom Jun Kim}
\affiliation{%
BK21 Physics Research Division and Department of Physics,
Sungkyunkwan University,Suwon 440-746,Korea
}%
\author{Bing-Hong Wang}
\affiliation{%
Research center for complex system science, Shanghai University for
Science and Technology, 200093, China
}%
\author{H. E. Stanley$^\dag$}
\affiliation{%
Center for Polymer Studies,Department of Physics,Boston
University,Boston,Massachusetts 02215,USA
}%

\date{\today}

\begin{abstract}
Motivated by the importance of geometric information in real
systems, a new model for long-range correlated percolation in
link-adding networks is proposed with the connecting probability
decaying with a power-law of the distance on the two-dimensional(2D)
plane. By overlapping it with Achlioptas process, it serves as a
gravity model which can be tuned to facilitate or inhibit the
network percolation in a generic view, cover a broad range of
thresholds. Moreover, it yields a set of new scaling relations. In
the present work, we develop an approach to determine critical
points for them by simulating the temporal evolutions of type-I,
type-II and type-III links(chosen from both inter-cluster links, an
intra-cluster link compared with an inter-cluster one, and both
intra-cluster ones, respectively) and corresponding average lengths.
Numerical results have revealed objective competition between
fractions, average lengths of three types of links, verified the
balance happened at critical points. The variation of decay
exponents $a$ or transmission radius $R$ always shifts the temporal
pace of the evolution, while the steady average lengths and the
fractions of links always keep unchanged just as the values in
Achlioptas process. Strategy with maximum gravity can keep steady
average length, while that with minimum one can surpass it. Without
the confinement of transmission range, $\bar{l} \to \infty$ in
thermodynamic limit, while $\bar{l}$ does not when with it. However,
both mechanisms support critical points. In two-dimensional free
space, the relevance of correlated percolation in link-adding
process is verified by validation of new scaling relations with
various exponent $a$, which violates the scaling law of Weinrib's.

\end{abstract}


\maketitle

\section{Introduction}
Correlated percolation\cite{c1, c2, c3, c4, c5, c6, c7, c8, c9, c10,
c11, c12, c13, c14} is a useful theoretic model in statistical
physics. It provides us with fundamental understanding of spread
processes of message, disease and matter in nature and society.
Linking probability between any two nodes in it takes the form of
$p(r)\sim r^{-a}$, where $r$ is $d$-dimensional distance between the
nodes, and $a$ is a positive real number, namely, distance-decay
exponent of links. Weinrib and Halperin\cite{WH} analytically
studied whether the correlations change the percolation behavior or
not. Weinrib\cite{Weinrib} pointed out, for $a<d$, the correlations
are relevant if $a\nu-2<0$, where $\nu$ is the percolation-length
exponent for uncorrelated percolation; while for $a>d$ the
correlations are relevant if $d\nu-2<0$. It is a generalization of
the Harris criterion\cite{Harris} appears earlier. Recently, network
models referring to correlated percolation have gradually
appeared\cite{c11}.

Achlioptas process(AP)\cite{Achlioptas} for link-adding networks,
which is an attractive topic at present\cite{2,3,4,5,6,7,
9,10,12,14,15,16,17,18,19,Dorogov, 2011 science, BJKim, grassberg,
tian}, could be viewed as a new kind of correlated percolation if we
put all nodes uniformly on a two-dimensional(2D) plane. Starting
from a set of isolated nodes, two candidate links are put to nodes
randomly at every time step, but only the link with smaller product
$m_i m_j$ is retained, where $m_i$(or $m_j$) is the mass (the number
of nodes) of the cluster that node i(or j) belongs to, which is
called Product Rule(PR). A link chosen with PR from both
inter-cluster candidates is called a type-I link. When two
candidate-links are of different types, i.e., one is an
inter-cluster link, the other is a intra-cluster one, always the
later is retained, and it is called a type-II link. While for both
intra-cluster ones, the retained link is arbitrarily chosen no
matter they are in the same or different clusters, and it is called
a type-III link. Generally speaking, in the way of AP, network
percolation is inhibited, which postpones the appearance of the
threshold $T_c$ at which a giant component G starts to grow, and
results in a sharp growth of G called an explosive percolation. In
our point of view, if we put AP on a 2D plane, it gives rise to a
new mechanism of long-range correlation for the nodes based on
co-evolutionary growing masses of components they connected. The
selective rule for topological links relies on mass-product instead
of 2D geometric length of them, which prevents the property
exhibited in the previous correlated percolation. And correlation
feature in AP-type of percolation has not been revealed up till now.

A recent model of ours\cite{me} based on the observation of
phenomena in different real systems\cite{percolation, MANET, global
connectivity, gravitation, Li, gravitation ad hoc} describes another
kind of correlated percolation in growing networks, which could be
viewed as a overlapping of traditional correlated percolation with
AP in a 2D space. The link-occupation function in the model takes
the form $p(r)\sim m_im_j/r_{ij}^a$ which looks like Newton's
gravity rule. It resumes the classical Erdos-Renyi(ER) random graph
model when exponent $a \rightarrow \infty$, and it gives another
extreme of AP when $a \rightarrow 0$. Different properties of such
kind of new correlated percolation are expected, since ER random
graph grows without any bias, AP takes a strong bias to inhibit
network percolation independent of geometric distance, while the new
model with gravity-like rule have some distance-related relax on
such bias, which produces a new type of correlated percolation.

In this paper, we report the simulation results on objective
competition between type-I, II and III links in both gravity
model\cite{gravitation, gravitation ad hoc} and AP model, and we
point out a new mechanism to support critical points, which bears
the scaling relations revealed in our recent work. Different
saturation effect is manifested, which distinguishes it from the
traditional correlated percolations.

\section{Model}
Suppose $N$ isolated nodes are uniformly scattered on a 2D plane.
For convenience of calculating distance, the plane is discretized
with a triangular lattice, each minimal edge with the length of two
units for the convenience of algorithm. Each vertex of the triangles
is occupied by a node so that we exclude all possible biases except
link-adding rules. For any two pairs of nodes $i$ and $j$ possibly
with the same product $m_i m_j$, a type-I link connects the the pair
with longer distance if both the links' ends hit the nodes belonging
to different clusters; while a type-II link connects
 the nodes inside the same cluster if the other one is an
 inter-cluster link; a type-III
 one connects arbitrarily chosen pair of nodes if both candidate
 links are intra-cluster
 ones.

Parallel to PR, we pick randomly two pairs [$(i,j)$ and $(k,l)$] of
nodes in the plane at every time step. For the pair $(i,j)$ (and for
$(k,l)$ likewise), we calculate the generalized gravity defined by
$g_{ij} \equiv m_i m_j/r^a_{ij}$, where $r_{ij}$ is the geometric
distance between $i$ and $j$, and $a$ is an adjustable decay
exponent. Once we have $g_{ij}$ and $g_{kl}$, we have two choices in
selecting which pair to connect. For the case of the maximum gravity
strategy (we call it $G_{\rm max}$) we connect the pair with the
larger value of the gravity, e.g., the link $(i,j)$ is made if
$g_{ij} > g_{kl}$ and the link $(k,l)$ otherwise. We also use the
minimum gravity strategy ($G_{\rm min}$) in which we favor the
smaller gravity pair to make connection. The two strategies, $G_{\rm
max}$ and $G_{\rm min}$, lead the link-adding networks to evolve
along the opposite paths of percolation processes. Generally
speaking, $G_{\rm max}$ facilitates the percolation process, whereas
$G_{\rm min}$ inhibits it. All such generalized gravity values are
calculated inside the circular transmission range with the radius
$R$ centered at one of nodes $i$ and $j$ as the speaking
node~\cite{Wangli2} in a mobile ad hoc network\cite{MANET, global
connectivity, gravitation ad hoc}. For the different limits of
parameters $R$ and $d$, we have three cases in the model. Case I:
With the transmission range $R\rightarrow \infty$, we have a
generalized gravitation rule which is an extension~\cite{gravitation
ad hoc} of widely used gravitation model~\cite{gravitation} $(a=1)$
with the tunable decaying exponent $a$. Case II: With the exponent
$a=0$, we assume that node pairs can be linked with PR topologically
inside the transmission range with a limited radius $R$. Case III:
With both limited values of radius $R$ and exponent $a$, we have the
gravity rule inside the transmission range. It can describe the
communication or traffics with constrained power or resources.

For case I and case III in the model, three scaling relations have
been found with
 large scale simulations. When strategy $G_{max}$ is adopted in 2D free space(case I), we have

\begin{equation}
 C\sim a^{-\alpha} f( t a^{\epsilon})
\end{equation}
 where $t=(T-T_0)/T_0$ is dimensionless time-step with $T_0=0.78$, $a$ is the decay
  exponent of connection probability.
$\alpha=0.01$, $\epsilon=0.20$, and $f(x)$ is a universal function.
When strategy $G_{min}$ is adopted inside the transmission range
with radius $R$(case III), we have
\begin{equation}
 C\sim (a/a_0)^{-\theta} h[t(a/a_0)^{\phi}]
\end{equation}
for certain parameter ranges of $a$ and $R$, where $t=(T-T_0)/T_0$,
$T_{0}=1.0$, $\theta=0.005$, $\phi=-0.50$, $a_0=0.5$, and $h(x)$ is
a universal function. In addition, when strategy $G_{max}$ for case
III is adopted inside transmission range defined by $R$, we have
another scaling relation
\begin{equation}
 C\sim R^{-\delta} H( t \rho^{\eta})
\end{equation}
for $R>3$, where $\rho=(R-R_0)/R_0$, $R_{0}=2$, $\eta=-0.10$,
$\delta=-0.005$, $T_0=1.0$ and $H(x)$ is a universal function.

To understand three scaling relations above, we should look into the
mechanism of the evolution processes underlying $C(T)$. To see what
happens in such critical points $T_{0}$, and what are particular of
them in certain link-adding processes, we count the temporal
 link fractions $F(T)$, and calculate the average lengths $\bar{l}$
 of links which is defined as
 the summation of all lengths of links for a certain type over its
 number in a window $\Delta T=20$ time-steps.
  By observation of the time-dependent behaviors
of fractions of type-I, II and III links, new properties were found
out for our gravity-like model together with AP producing explosive
percolations.

\section{3. Simulation results}

All simulations are carried out on the triangular lattice of the
size $N = L\times L$ with $L=32, 64, 128$ and $256$, respectively.
We simulate either of strategy $G_{\rm max}$ or $G_{\rm min}$ for
either case I or III. The total number of links equating to that of
time-steps is divided by $N$, which is defined as $T$. The mass of
the largest component divided by $N$ makes up the observable $C$,
the node fraction of the largest component. All results presented in
this work are obtained from 5000 different realizations of network
configurations with $L=128$ if not specially indicated.

\begin{figure}
\includegraphics[width=3.5in]{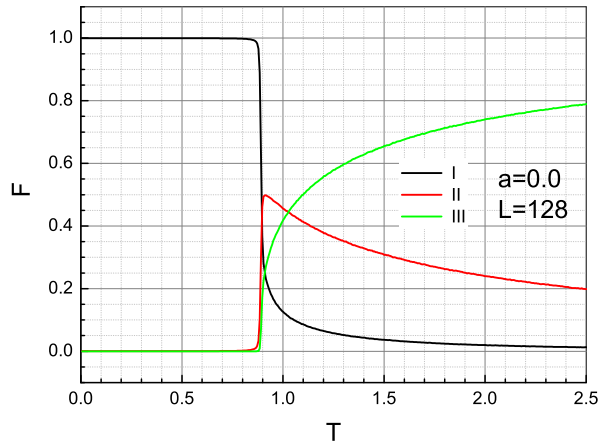}
\caption{(color online)Evolution of fractions of
type-I, type-II and type-III links in Achlioptas process on 2D
plane.}
\end{figure}

\begin{figure}
\includegraphics[width=3.5in]{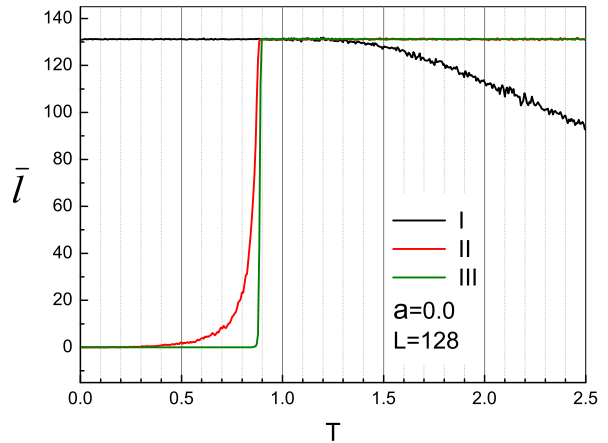}
\caption{(color online)Evolution of average
lengths of type-I, type-II and type-III links in Achlioptas process
on 2D plane.}
\end{figure}

Inspired by Cho and Kahng's \cite{19} work and a referee of
ref.\cite{me}, we have gone further by calculating fractions of
three types of links and arithmetic average lengths of their links.
Our attention was pointed at AP first. In Fig.1 we illustrated the
evolution of fractions of 3 types of links. Just at the threshold
the fraction of type-I$(F_{I})$ links has a sharp drop-down,
meanwhile that of type-II $(F_{II})$ shoots up, crossing $F_I$ at
$T_c=0.888$. A little after it, $F_I$ crosses with growing fraction
of type-III links ($F_{III}$) at the level $F_I=F_{III}=0.25$, while
$F_{II}$ gets its summit($F_{II}=0.5$) at the same point, which has
not been concerned by previous works. However, it is this property
that pervades all cases in the present correlated percolation. In
Fig.2, the average lengths of type-II($\bar{l}_{II}$) merges that of
type-III($\bar{l}_{III}$) at $T_c$ after an abrupt growth,
$\bar{l}_{II}$ starts to grow earlier than $\bar{l}_{III}$. The
level of $\bar{l}$ for both of them keep invariant for $T>T_{c}$,
while $\bar{l}_{I}$ starts to decrease from $T_c$. We see from both
the figures that in explosive percolation the system undergoes a
sharp transition from a type-I link dominant phase into a type-II
and III dominant phase at $T_c$\cite{19}. Besides, average lengths
undergo a parallel transition at the same point. Actually, 3 levels
of $\bar{l}$ go to infinity in dynamic limit from finite size
scaling transformation(not shown).

\begin{figure}
\includegraphics[width=3.5in]{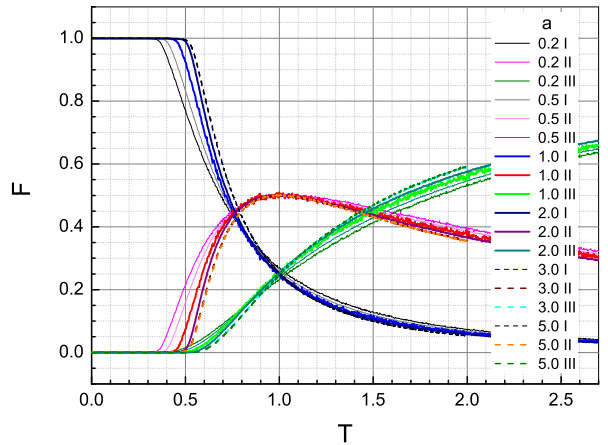}
\caption{(color online)Evolution of
fractions of type-I, type-II and type-III links with $G_{max}$ and
probability decay exponent $a=0.2,0.5, 1.0, 2.0, 3.0 and 5.0$ in
case I ($R\rightarrow \infty$).}

\end{figure}

\begin{figure}
\includegraphics[width=3.5in]{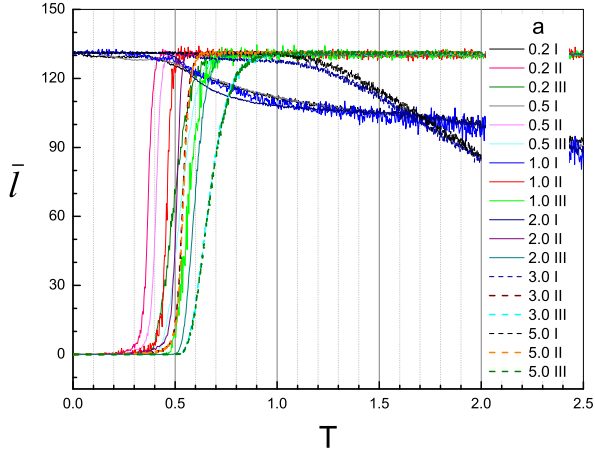}
\caption{(color online) Evolution of
average lengths $\bar{l}$ of type-I, type-II and type-III links with
$G_{max}$ and probability decay exponent $a=0.2,0.5, 1.0, 2.0, 3.0
and 5.0$ in case I.}

\end{figure}

\begin{figure}
\includegraphics[width=3.5in]{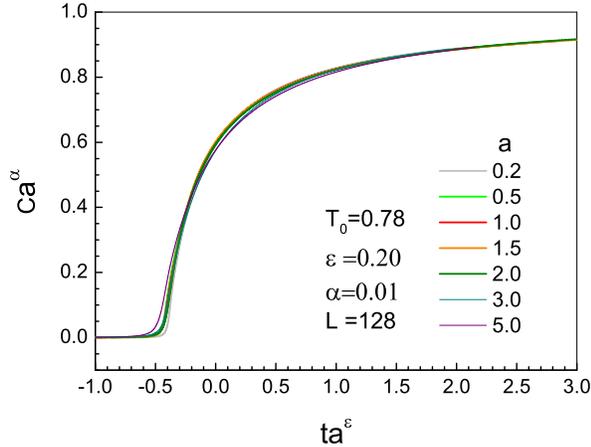}
\caption{(color online) Node fraction
$C(T)$ of the largest component with strategy $G_{\rm max}$ and the
same parameters in Fig.3.}

\end{figure}

Now we turn to the fraction of 3 types of links in case I of the
 present gravity model\cite{me}. With
strategy $G_{max}$ in free 2D space, we have scaling relation (1)
for distance-decay exponent $a \in[0.2, 2.0]$. Fig.3 shows the
evolution of fractions corresponding to it. Curves for $F_I$ cross
those of $F_{II}$ with $a=0.5, 1.0$ and $2.0$ around $T=0.78$ quite
clearly, even those with $a=0.2$ barely cross near it. But fractions
for $a=3.0$ and $5.0$ shift the cross point rightward obviously. On
the other hand, the curves for $F_{I}$ cross $F_{III}$ at $T=1.0$
for almost all values $a$ except $a=0.2$. To determine which one
would be the candidate of another critical point, we cast ourselves
on the assistance from the observation of average lengths of links.
In Fig.4 $\bar{l}_{II}(T)$ merges $\bar{l}_{II}(T)$  for $a=3.0$ and
$5.0$ at $T=1.0$, separating themselves from others. However, almost
all other ones collect at $T=0.78$, which gives hint to us for
$T_0$.(Here a better resolution is needed in further calculation).
Correspondingly, in Fig.5 for scaling relation (1) with exponents
$\epsilon=0.2$, $\alpha=0.01$, and $T_{0}=0.78$,
 $C(T)$ for $\alpha \in [0.5, 2.0]$ collapse into the universal
 function very well, with that for $a=0.2$ barely collapsing onto
 it. But those $C(T)$ for $a=3.0$ and $5.0$ do not behave well
 in collapse. The separation from others at the turning middle
 part indicates the deviation of their $T_0$
 from $0.78 $ with which others share. In the description of
 the average lengths for case I with $G_{max}$, simulated results
 $\bar{l}(T)$ in Fig.4 with all values $a$ demonstrate the same
 steady level ($\bar{l}\simeq 131.50$ for $L=128$). Variation
 of parameter $a$ only shifts starting points of up-growing
 $F_{II}$ and $F_{III}$ as $a$ increases. The saturation effect
 of large decay exponents $(a=3.0, 5.0)$ appears clearly and
 is shown by dash lines, which demonstrates the inheritance from
 traditional correlated percolation\cite{Weinrib}. In this case
 with $G_{max}$, special level of fractions at cross point
 of $F_{I}(T)$ and $F_{III}(T)$ keeps 0.25 just as in AP without
 any distance-decay included, so does $F_{II}$ at hiking its summit.

\begin{figure}
\includegraphics[width=3.5in]{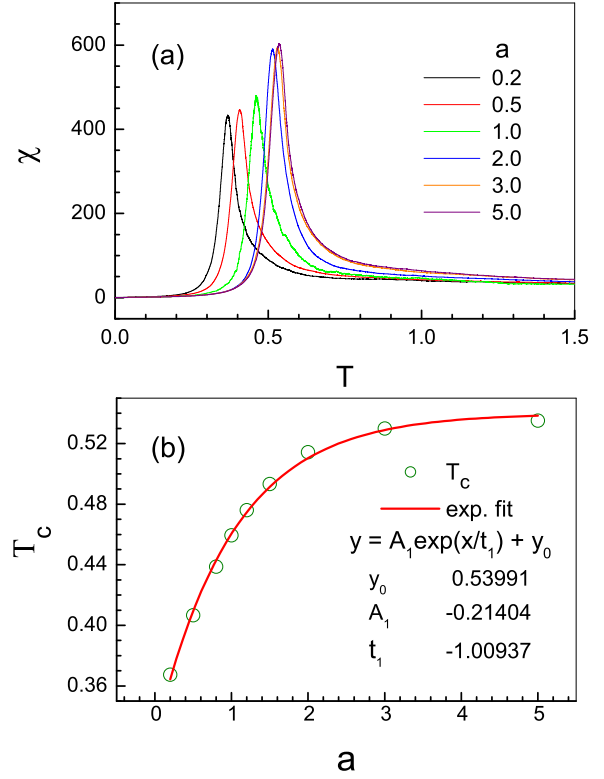}
\caption{(color online)(a)Susceptibility
$\chi(T)$ of the system with $G_{max}$ in Case I. $L=32, 64, 128$
and $256$.(b)Percolation thresholds $T_c$ for $a=0.2, 0.5, 0.8, 1.2,
2.0, 3.0$ and $5.0$ with $G_{max}$ in Case I.}
\end{figure}

\begin{figure}
\includegraphics[width=3.5in]{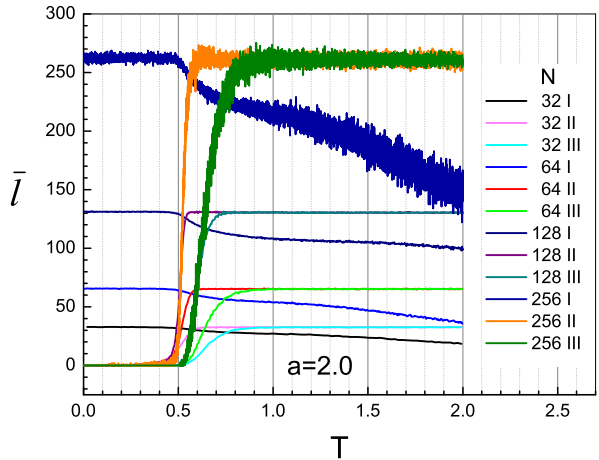}
\caption{(color online)Size-dependent
average lengths $\bar{l}$ of type-I, type-II and type-III links with
strategy $G_{\rm max}$ in case I. Parameters are the same as in
Fig.6}

\end{figure}

As in a usual way, we determine the critical point $T_c$ of
percolation by observation of tips of susceptibility $\chi(T)$
(Fig.6). Comparing $T_c$ in Fig.6 with $\bar{l}(T)$ in Fig.4, we
find that these $T_c$ approximately hit the horizontal coordinates
of middle point of growing fraction of type-II links, which means
that $T_c$ is the transition point from the inter-cluster-link
dominant phase to the intra-cluster-link dominant phase. Besides
$T_c$, we have another (sub)-critical point $T_0$ which is in
certain range independent of decay exponent $a$ in gravity model,
and $T_0$ indicates the balance between the fractions of type-I and
type-II links, yielding a new scaling behavior of $C(T,a)$ in
formula (1) not revealed by previous works. Moreover, the steady
level $\bar{l} = 131.5$ is always $a-$independent, type-independent
which takes the inherited value of that
 in AP. Actually, 131.5 is the value for L=128 only. We have size
 effect since a free boundary condition instead of a periodic one
 is adopted. The finite size effect is shown in Fig.7 which gives that
$\bar{l}\sim L $ , i.e.,

\begin{equation}
\bar{l} \sim N^{1/2}
\end{equation}

where $N$ is the number of nodes on the 2D plane. Hopefully it goes
towards infinity in thermodynamic limit. However, the finite size
effect of $C(T,N)$ (Fig.8 for an example) is not strong enough for
us to identify the scaling exponents $\nu, \beta$ as usual.
Therefore, we can check the validation of scaling laws presented by
Weinrib \cite{Weinrib} for correlated percolation in the present
model only by rescaling susceptibility $\chi (a,T)$. Fig. 9
illustrates the results of it for examples $a=0.5$ and $2.0$,
respectively. We have scaling relation

\begin{figure}
\includegraphics[width=3.5in]{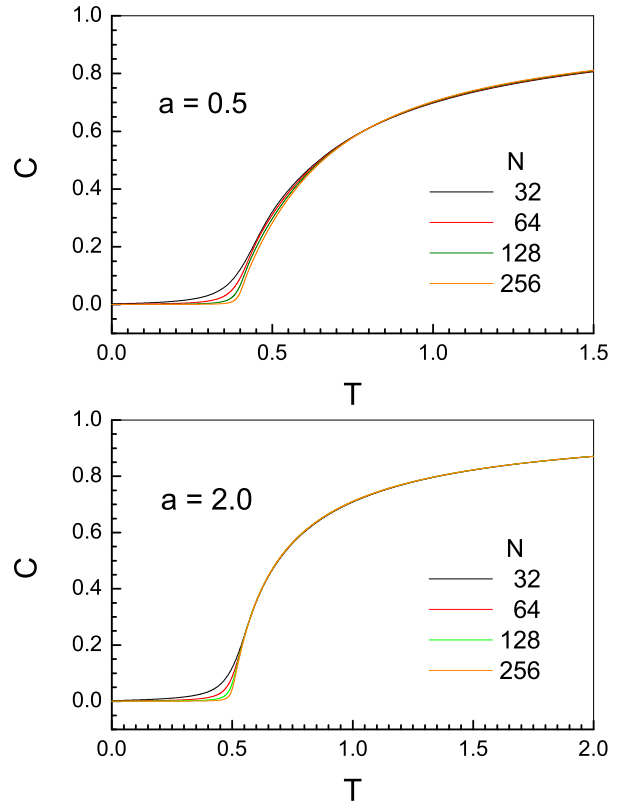}
\caption{(color online)Node fraction
$C(T)$ of the largest component with strategy $G_{\rm max}$ in Case
I. (a)$a=0.5$; (b)$a=2.0$. $N=32, 64, 128$ and $256$.}
\end{figure}

\begin{figure}
\includegraphics[width=3.5in]{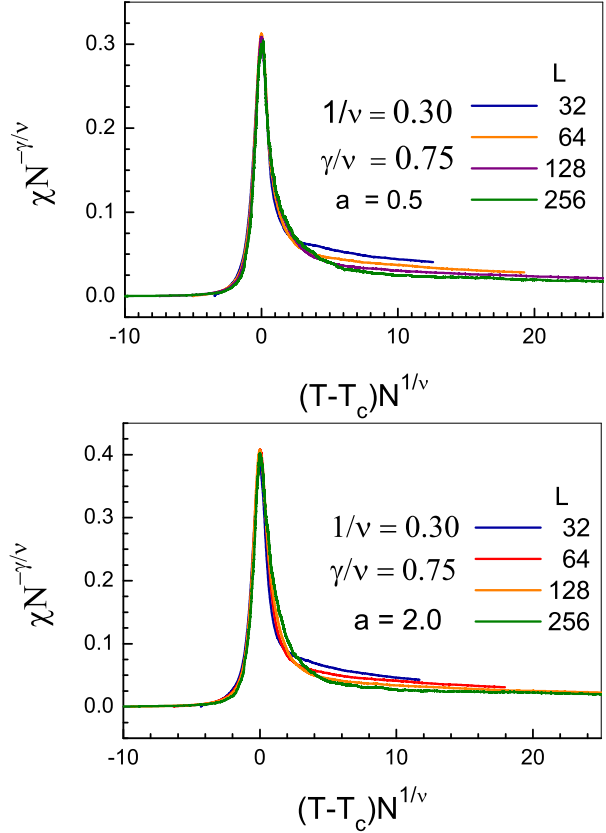}
\caption{(color online) Scaling of
susceptibility $\chi(T)$ in case I for (a) $a=0.5$; (b) $a=2.0$.
$N=32, 64, 128$ and $256$.}
\end{figure}

\begin{equation}
\chi \sim N^{\gamma/\nu}G((T-T_c)N^{1/\nu})
\end{equation}
where $1/\nu=0.3$, $\gamma/\nu=0.75$, and $G(x)$ is a universal
function. With these values of scaling exponents, the scaling law
$\nu_{long}=2/a$ is not applicable to the present model, where
$\nu_{long}$ is correlation-length exponent in the long-range case.

With power-law form for the correlation function g(r), Weinrib had
derived the extended Harris criterion: the long-range nature of the
correlations is relevant if $a\nu -2 < 0$, which means the
correlations change the percolation critical behavior. It has been
violated since now they all behave differently from traditional
short range percolation in a 2D triangular lattice ($\nu =4/3$) and
the correlations are relevant no matter $a \nu-2$ is less(Fig.9(a))
or larger(Fig.9b) than zero. This is because in strategy $G_{max}$
we have overlapped the power-law correlation function g(r)
 with AP which is another kind of autocorrelation process with positive
 feed back effect of mass-growing.

\begin{figure}
\includegraphics[width=3.5in]{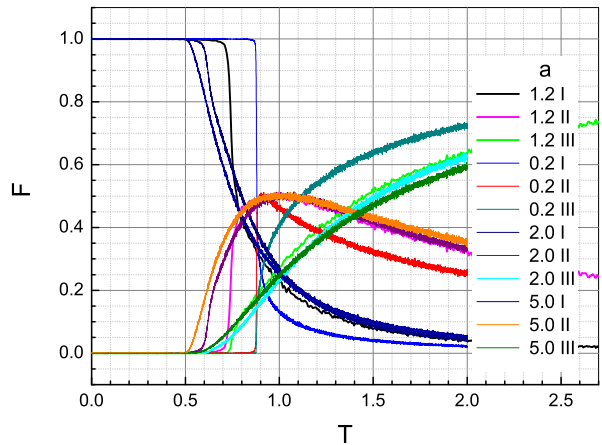}
\caption{(color online)Evolution of
fractions of type-I, type-II and type-III links with $G_{min}$ and
probability decay exponent $a=0.2,1.2,2.0$ and $5.0$ in case I.
$L=256$.}

\end{figure}

\begin{figure}
\includegraphics[width=3.5in]{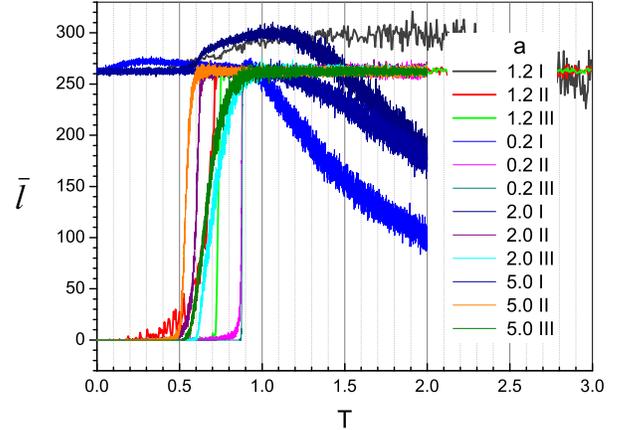}
\caption{(color online)Evolution of
average lengths of type-I, type-II and type-III links with $G_{min}$
and probability decay exponent $a=0.2,1.2,2.0$ and $5.0$ in case I.
$L=256$.}
\end{figure}

For the strategy $G_{min}$ which prefers smaller gravity, it tends
to retain a longer link under the comparison of the same product of
masses $m_{i}m_{j} $. That is to say, long range links have
predominance. Evolution of $F_{I}(T)$ and $F_{II}(T)$ links do not
cross at any
 common point. In Fig.10, the cross points for $F_{I}$ and $F_{II}$
 shift leftward from $T_c$
 of AP as decay exponent $a$ increases, which means that $G_{min}$ as
 a correlated
 percolation mechanism
 weaken the explosive effect caused by AP. But the
 starting position of fraction-II
 still provides hints of thresholds $T_c$. However, it is hard to
 locate a common cross point
  for $F_{I}$ and $F_{III}$ in a range of $a$. Therefore, we have
  no scaling relation for them.
   The steady average lengths of links (Fig.11) have the same level
   of $G_{max}$ and AP cases,
    and size-effect (not shown) also tells the divergence of $\bar{l}$,
    but all of them do not
     tell any possible hint for critical points.

Generally speaking, strategy $G_{min}$ facilitates longer links for
certain geometric distribution of clusters or nodes. In the
evolution, strategy $G_{min}$ emphasis the assignments for
 different types of links,
 encourage longer and intra-cluster links. Humps above the steady
 $F_{II}$ and $F_{III}$
 implies out-of-pace growing of link lengths of $F_{I}$. That
 is, $F_{I}$ surpasses the
 growing speed of the giant component. Here, it is the geometric
 distance-dependent strategy
 that makes $G_{min}$ alleviate effect of AP. With smaller $a$
 (e.g., a=0.2 in Fig.11)the
 strategy has the opportunity to exhaust long links before $T_c$;
 while with middle values
  of $a$ (e.g., $a=1.2$ and $2.0$) it may take longer time to exhaust
  them. However, with too
  large $a$ (e.g., $a=5.0$), $G_{min}$ fall off quicker than the
  natural dimension, we can
  only see the AP-type short-range effect of saturation. In this
  limit, i.e., $a\leq 3.0$,
  percolations are no longer relevant, which causes saturation
  of curves $C(T)$ in
  Fig.1b \cite{me}. However, we should not expect the short-range
  percolation exponent
  $\nu=4/3$ of correlation length here for 2D triangular lattice,
  since AP has been
  included in the strategy $G_{min}$.

\begin{figure}
\includegraphics[width=3.5in]{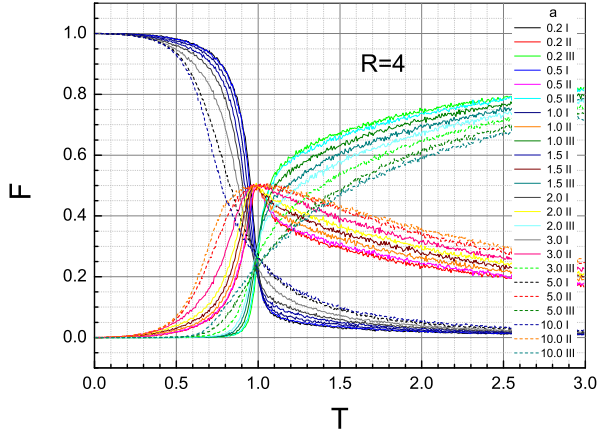}
\caption{(color online)Evolution of
fractions of type-I, type-II and type-III links with strategy
$G_{min}$ and probability decay exponent $a=0.2, 0.5, 1.0, 1.5, 2.0,
3.0, 5.0$ and $10.0$ in case III. $R=4$.}
\end{figure}

\begin{figure}
\includegraphics[width=3.5in]{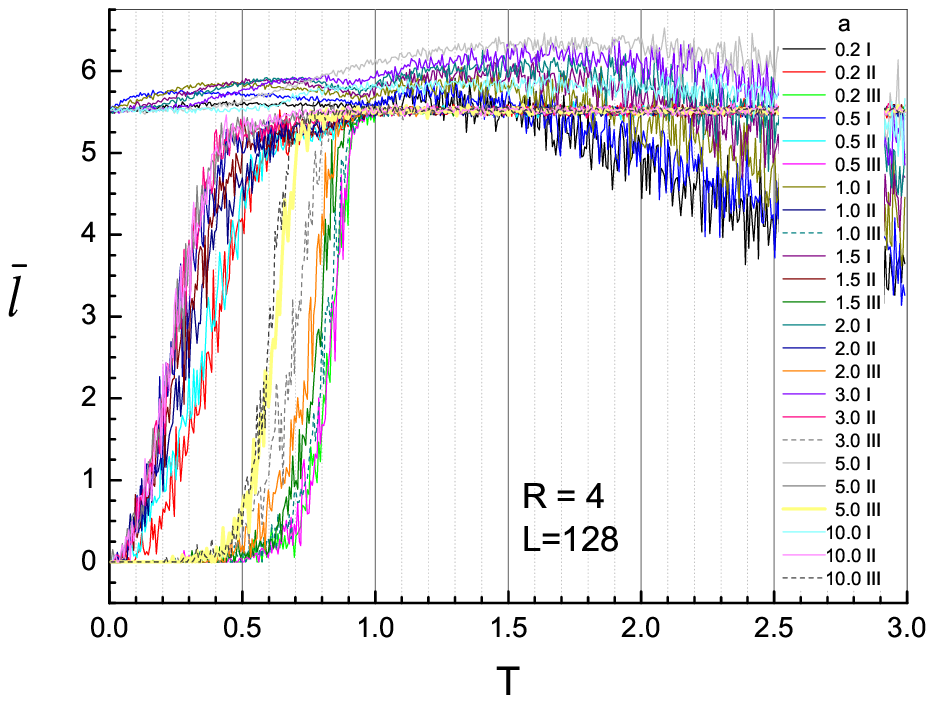}
\caption{(color online)Evolution of
average lengths of type-I, type-II and type-III links with strategy
$G_{min}$ and probability decay exponent $a=0.2, 0.5, 1.0, 1.5, 2.0,
3.0, 5.0$ and $10.0$ in case III. $R=4$.}
\end{figure}

\begin{figure}
\includegraphics[width=3.5in]{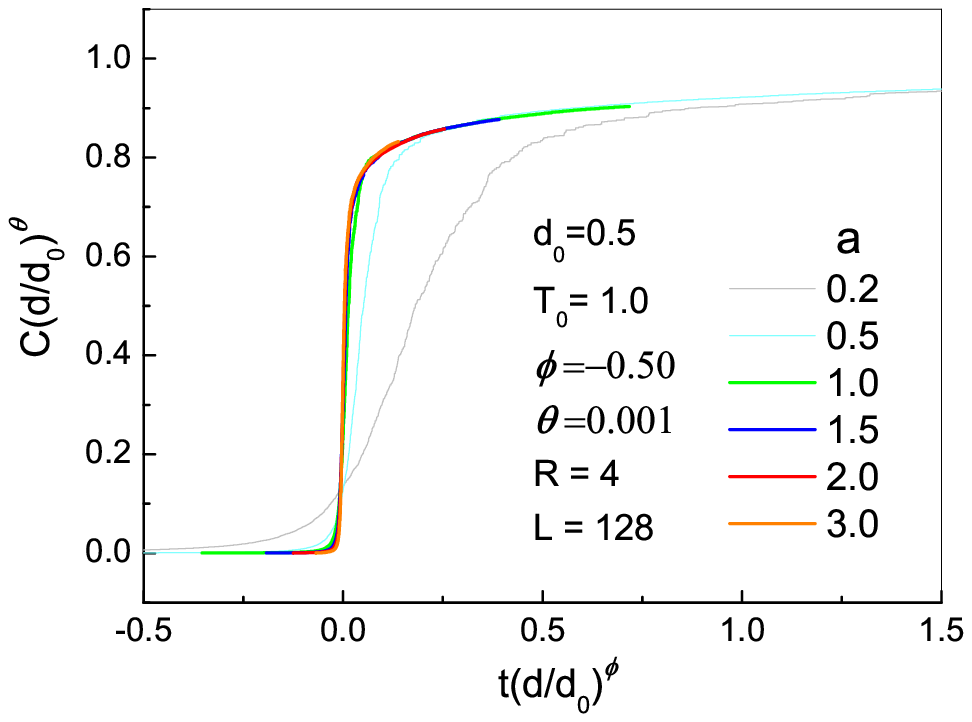}
\caption{(color online)Scaling of node
fractions $C(T)$ with strategy $G_{min}$, $a=1.0, 2.0$ and $3.0$in
Case III. $R=4$.}
\end{figure}

\begin{figure}
\includegraphics[width=3.5in]{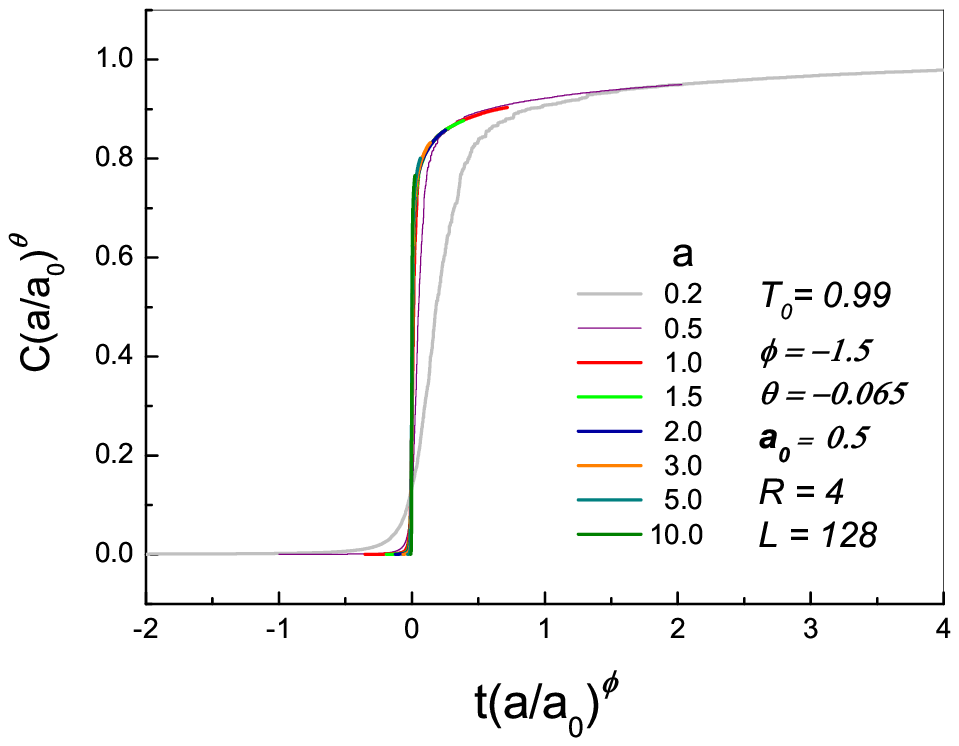}
\caption{(color online)Pseudo-scaling of
node fractions $C(T)$ with strategy $G_{min}$, $a=0.2, 0.5, 1.0,
1.5, 2.0, 3.0, 5.0$ and $10.0$ in Case III. $R=4$.}
\end{figure}

\begin{figure}
\includegraphics[width=3.5in]{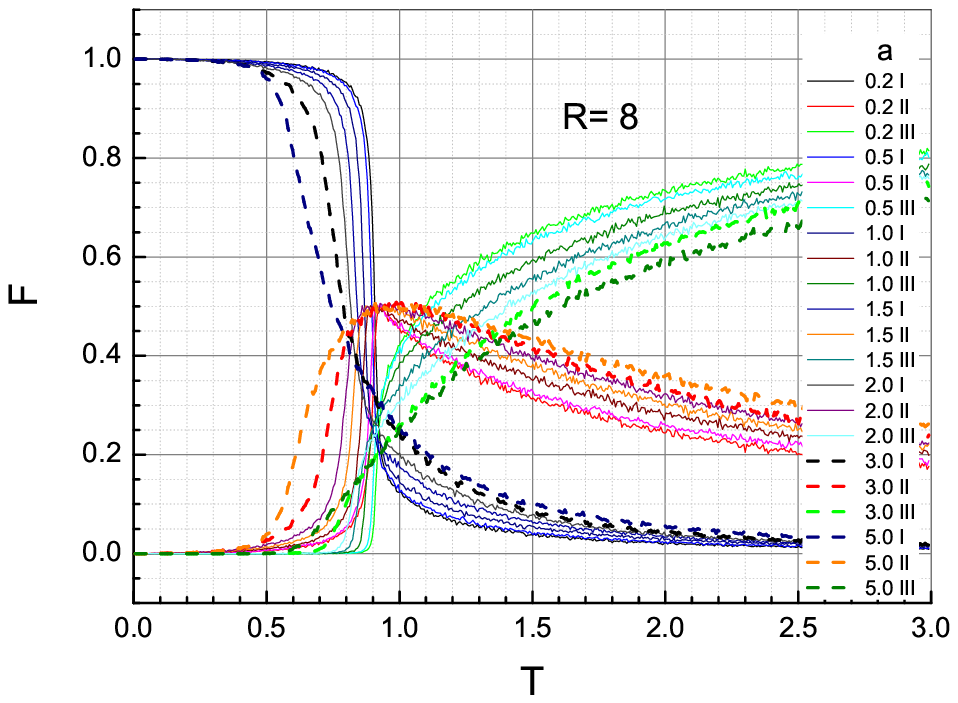}
\caption{(color online)Evolution of
fractions of type-I, type-II and type-III links with strategy
$G_{min}$ and probability decay exponent $a=0.2, 0.5, 1.0, 1.5, 2.0,
3.0$ and $5.0$ in case III. $R=8$.}
\end{figure}

The distinct feature in case III for both $G_{min}$ and $G_{max}$ is
that the candidate links are selected not only by comparing
gravities, but also constrained inside a transmission range $R$
($r=2R$ in geometric distance), which ruins the effect comes from
the divergence of average link lengths. It is well known that all
possible singularities at critical points come from the singularity
of correlation length. However, here no length could goes to
infinity in any way, which ruins possible common cross point relies
on the balance between $F_{I}$ and $F_{II}$ as in the 2D free space
(case I of the present model)and induces the possibility to yield
novel scaling relation other than any previous ones. For possible
critical point, we seek help from the evolution of link fractions of
3 types. Fig.12 shows the behaviors of $F_{I}$, $F_{II}$ and
$F_{III}$ with $G_{min}$ for all simulated distance-decay exponents
$a$ and $R=4$.

Correspondingly, Fig.13 shows the behaviors of $\bar{l}_{I}$,
$\bar{l}_{II}$ and $\bar{l}_{III}$ the the same set of parameters.
Critical point $T_{0}=1.0$ ($T_{0}=0.99$ to be precise)
distinguishes itself from others
 by intuitive observation. The cross point for $a=5.0$ and $a=10.0$
 go rightward from which others share, which means
 they could not share the same $T_{0}$ hence the same scaling
 relation with other decay exponents. Besides,
 in rescaling process for $C(T,a)$, curves for $a=0.2$ and $a=0.5$
 failed in collapse, because smaller transmission
 range $R$ inhibits the effect of slower(long-range) decay for
 connection probability. The rescaled function $C(t)$
 for $R=4$ is shown in Fig.14. It seems that we could go
 further with $\phi=-1.5$ to include more exponents $a$ in
 the scaling as shown in Fig.15. However, it is meaningless
 in physics due to above mentioned reasons. Actually, scaling
  behaviors are R-dependent, but exponents $\phi$ and $\theta$
  need not to vary. The variation of $R$ only shifts
  $T_{0}$ as illustrated in Fig.16 for $R=8$. Therefore, we
  keep $\phi=-0.5$, $\theta=0.005$ for all values of $R$
  with $G_{min}$, but take $T_{0}=0.99$ for $R=4$, $T_{0}=0.92$
  for $R=8$, and so on. The humps above the steady level of
  $\bar{l}_{II}$ and $\bar{l}_{III}$(all independent of
  exponents $a$) come from similar mechanism as in free 2D
  space but now at much lower level constrained by transmission
  radius $R$, and they are independent of size $L$ of the system.

\begin{figure}
\includegraphics[width=3.5in]{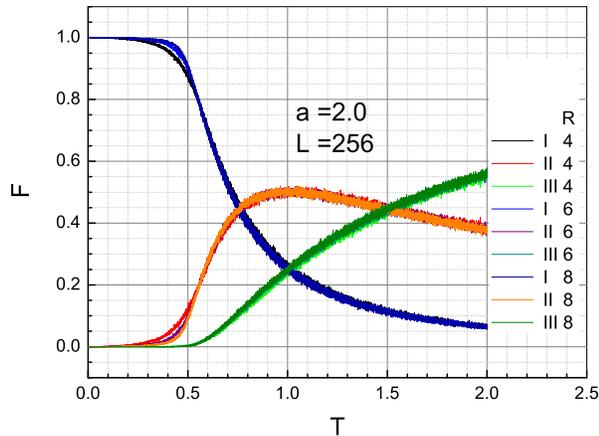}
\caption{(color online)Evolution of fractions of type-I, type-II and
type-III links with strategy $G_{max}$ and probability decay
exponent $a=2.0$ for $R=4,6$ and $8$ in case III. $L=256$. 500
realizations of network configurations.}

\end{figure}

\begin{figure}
\includegraphics[width=3.5in]{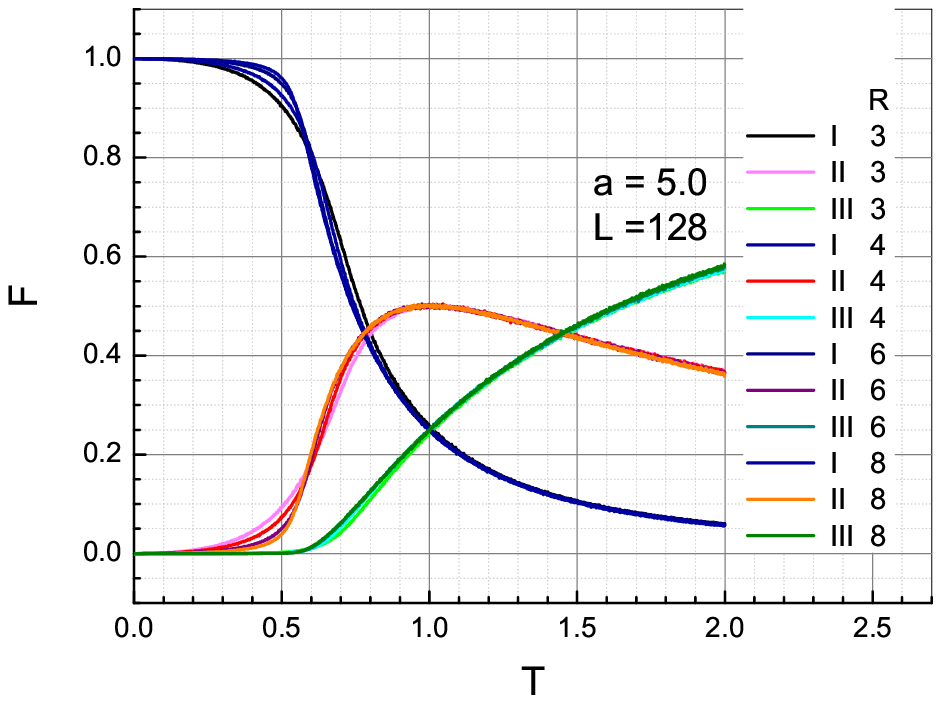}
\caption{(color online)Evolution of
fractions of type-I, type-II and type-III links with strategy
$G_{max}$ and probability decay exponent $a=5.0$ for $R=3, 4, 6$ and
$8$ in case III.}
\end{figure}

\begin{figure}
\includegraphics[width=3.5in]{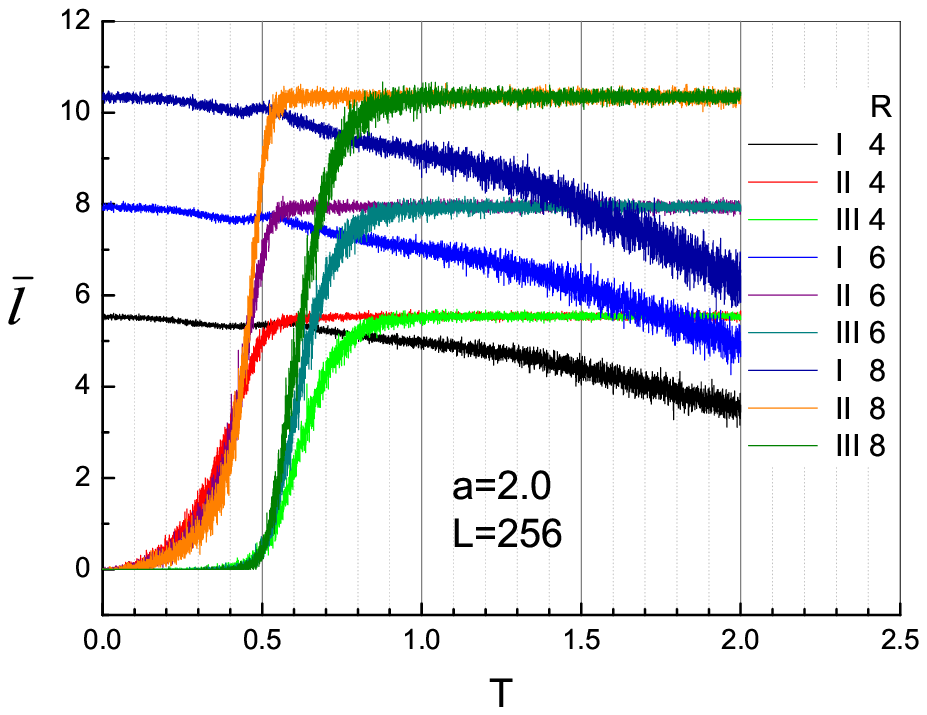}
\caption{(color online)Evolution of
average lengths of type-I, type-II and type-III links with strategy
$G_{max}$ and probability decay exponent $a=2.0$ for $R=4,6$ and $8$
in case III. $L=256$. 500 realizations of network configurations.}
\end{figure}

\begin{figure}
\includegraphics[width=3.5in]{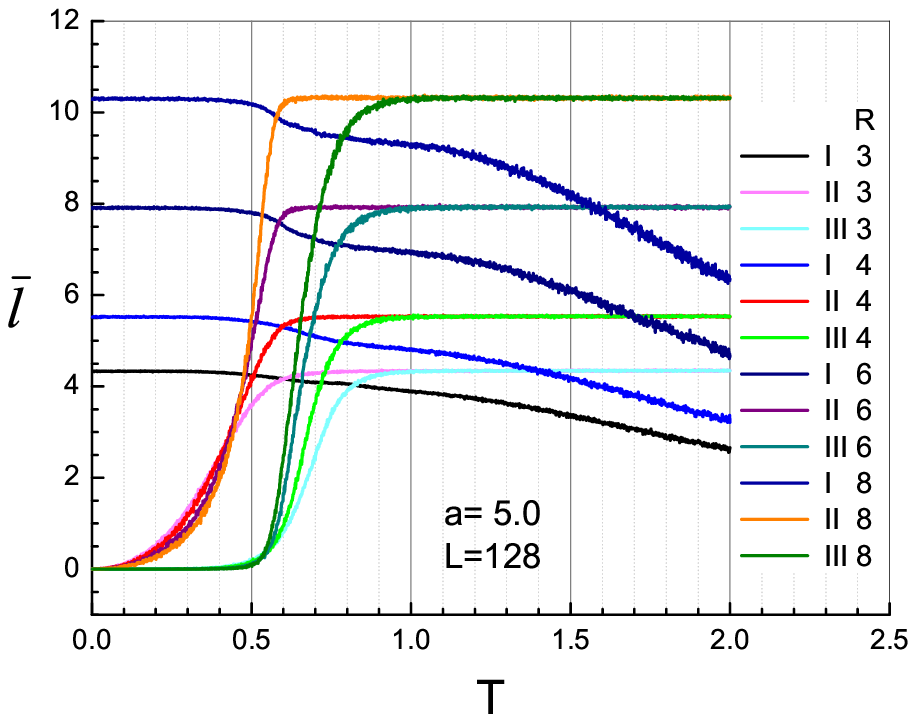}
\caption{(color online)Evolution of
average lengths of type-I, type-II and type-III links with strategy
$G_{max}$ and probability decay exponent $a=5.0$ for $R=3, 4, 6$ and
$8$ in case III.}
\end{figure}

\begin{figure}
\includegraphics[width=3.5in]{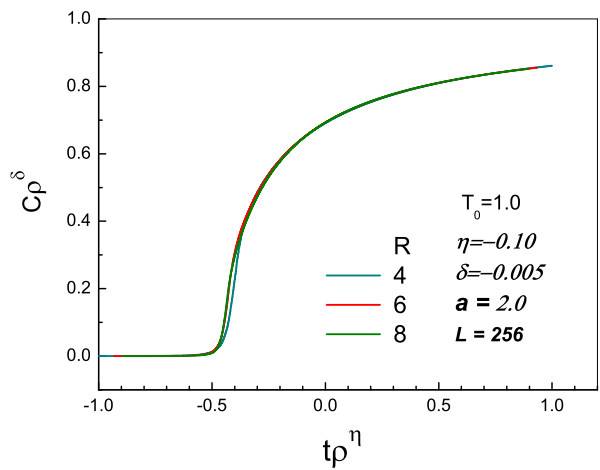}
\caption{(color online)Scaling of node
fractions $C(T)$ with strategy $G_{max}$, $a=2.0$  for $R=4, 6$ and
$8$ in Case III.$L=256$. 500 realizations of network
configurations.}
\end{figure}

\begin{figure}
\includegraphics[width=3.5in]{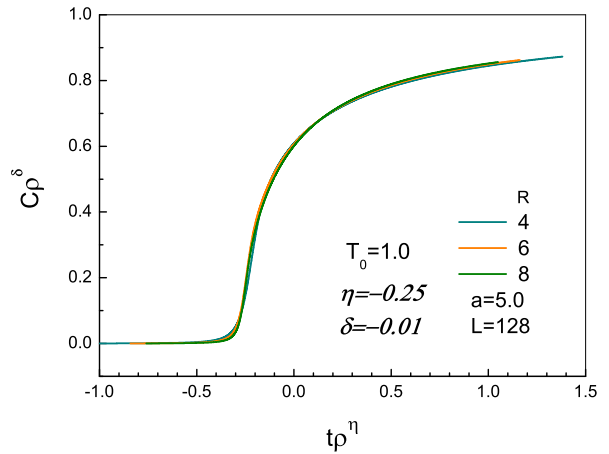}
\caption{(color online)Scaling of node
fractions $C(T)$ with strategy $G_{max}$, $a=5.0$  for $R=3, 4, 6$
and $8$ in Case III.}
\end{figure}

  Scaling relation (3) for case III with strategy $G_{max}$ inside transmission range
  with radius $R$ is checked for various decay exponents $a$ and for different sizes
  (L=32, 64, 128 and 256). Its validation is independent of size $L$ simulated. In Fig.17
  and Fig.18, the evolution of $F_{I}$, $F_{II}$ and $F_{III}$ for both $a=2.0$ and
  $a=5.0$ behave much similarly. $F_{I}$ and $F_{II}$ cross at the level a little bit lower
  than $0.45$, while $F_{I}$ crosses
  $F_{III}$ at the level $0.25$, which keeps the same as in all previous cases. The
  changes of exponent $a$ and $R$
  only shift fractions along horizontal direction of figures, i.e., to change starting
  points and growing/dropping speed instead of levels of them. However, $T_0=1.0$ keeps
  as their common fixed point for $F_{I}$ and $F_{III}$ to cross. In Fig.19 and Fig.20,
  $F_{I}$ and $F_{III}$ for both $a=2.0$ and $a=5.0$ arrive at the same level hitting
  $T_{0}=1.0$, which distinguishes this point from totally 3 cross points, and makes
  up a candidate of critical point $T_0$ for scaling relations. The scaling exponents
  $\eta=-0.10$ and $\delta=-0.005$ have been checked for lower values of parameter
  $a$($0.5\leq a \leq 3.0$, Fig.21). However, for $a=5.0$, we have to choose a new set
  of exponents: $\eta=-0.25$ and $\delta=-0.01$(Fig.22). It is not strange that steady
  levels of $\bar{l}$ keep unchanged for certain $R$, independent of $a$ or $L$, just as
  that with $G_{min}$ in case III. $\bar{l}$ inside a circle defined by $R$ can not
  go
  to infinity under any circumstance, but still support a critical point,
  which distinguishes correlated percolation in
  case III from case I and traditional models. It deserves further
  investigation.

\section{Discussion and Conclusions}
In this paper, we have proposed a new network model of correlated
percolation in which geometric distance-dependent power-law decay
connection probability overlaps Achlioptas process to form a gravity
model. It can be tuned to facilitate or inhibit percolation with
strategy $G_{max}$ or $G_{min}$, cover a wide range of thresholds
$T_c$, yield a set of new scaling relations. And it provides a
scheme for better description of practical processes in complex
systems.

We have developed a new approach to find out candidate critical
points with physical meanings other than that of traditional ones.
There are objective competition and balance between type-I and
type-II, type-I and type-III links, meanwhile, competition of
average lengths between type-II and type-III links. Along this line
threshold $T_c$ is found to overlap the balance point between
factions $F_I$ and $F_{II}$ in the explosive percolation of
Achlioptas process, and the steady average lengths of three types of
links are all divergent to infinity in thermodynamic limit. The
percolation is indeed a transition from type-I link dominant phase
to type-II and type-III dominant phase.

By observing evolutions of fractions of type-I, type-II and type-III
links, a candidate critical point can be chosen combined with the
message on evolutions of average lengths of them. With strategy
$G_{max}$ in 2D triangular lattice, fraction $F_{I}$ get balance
with $F_{II}$, makes up a critical point $T_0$ which supports
scaling relation (1) in case I of the model. With strategy $G_{min}$
inside certain transmission range with radius $R$, a duet balance
exists for $F_{I}$ and $F_{III}$ meanwhile $\bar{l}_{II}$ and
$\bar{l}_{III}$, makes up another critical point $T_0$ which
supports scaling relation (2) in case III. With strategy $G_{min}$
and certain range $0.5 \leq a \leq 3.0$ of decay exponent $a$, again
a duet balance exists for $F_{I}$ and $F_{III}$ meanwhile
$\bar{l}_{II}$ and $\bar{l}_{III}$,
 makes up another critical point $T_0$ which supports scaling relation
 (3) for a mini-scale
of $R$ in case III. This approach serves an assistant tool in
seeking critical points of order parameter $C(T)$ which is usually
not easy to determine in an intuitive way.

In numerical calculations, besides percolation threshold $T_c$, two
fixed points, $T_0=0.78$ and $T_0=1.0$ emerge as distinct points not
only for special temporal crux but also for unchanged levels of
$F_I$, $F_{II}$ and $F_{III}$ inherited from AP, which is expected
to be further proved in analytical ways. However, they have
different physical meanings. The former corresponds to a divergent
average length of links, while the later corresponds to confined
average lengths by transmission range $R$, which distinguishes
itself from traditional critical points in percolations.

Correlated percolations are relevant since long-range correlation
drastically changes the critical properties. The validation ranges
of decay exponents $a$ with various strategies in different cases
define the relevance of correlation. They have demonstrated novel
scaling relations different from traditional 2D short-range
percolation in triangular lattice. The intervention of
distance-dependent power-law decay ingredients alleviates the
explosive effect of percolation transition by horizontal adjustment
of evolutions along temporal axis, separates $T_0$ from $T_c$, while
the overlapped AP included in the present gravity model always
conquers the vertical levels of three fractions and average lengths,
which are
 found neither
in traditional correlated percolations of continuities in 2D space
nor in complex networks. Moreover, the node fraction $C(T)$ of the
largest component, fractions $F(T)$, and average lengths
$\bar{l}(T)$ of three types of links all show saturation phenomena
as pointed out by Weinrib but with different values of exponent of
$a$ since now AP overlaps in the present gravity model. And scaling
law of Weinrib is no longer obeyed according to the evidence of
numerical results of average-length exponents.

\begin{acknowledgments}
We are indebt to anonymous referees for stimulating comments. Zhu
thanks H. Park, P. Holm, X.-S. Chen and Z.-M. Gu for useful
discussion. We acknowledge financial support from National Natural
Science Foundation of China (NNSFC) under Grants No. 11175086,
10775071 and 10635040. Kim was supported by the National Research
Foundation of Korea (NRF) funded by the Korea government (MEST)
under Grant No. 2011-0015731.
\end{acknowledgments}

{}

\end{document}